# Ultrafast Magnetic Resonance Imaging

Michael Hutchinson, [1] Ulrich Raff [2,3] and Luis Osorio [3]

1. Department of Neurology, Icahn School of Medicine, One Gustave Levy Place, New York NY 10029
2. Department of Physics, University of Santiago, Santiago, Chile.
3. Department of Electrical Engineering, University of Santiago, Santiago Chile.

Correspondence to: Michael Hutchinson, M.D., Ph.D., email:michael.hutchinson@mssm.edu

## Abstract

We propose a radical advance in Magnetic Resonance Imaging, which, after almost half a century, remains slow. This is entirely because MRI requires successive applications of magnetic field gradients to encode for spatial location, and why it is still largely restricted to imaging of the brain and joints, body parts that do not move with respiration. Parallel MRI accelerates imaging by permitting undersampling of k-space using multiple detectors, each with a unique spatial sensitivity to the radio field emitted by the object. This is called spatial sensitivity encoding. A fourfold undersampling means a fourfold acceleration, however a five-fold undersampling cannot be reconstructed. This is because too much reliance is placed on spatial sensitivity encoding, and spatial sensitivity encoding is less efficient than gradient encoding when the detectors are large.

On the contrary, we hypothesized that when very large numbers of *very small* detectors are deployed, it would render spatial sensitivity encoding efficient, and would allow *complete* elimination of gradient reversals and therefore extremely rapid MR imaging. Since each detector receives all of the signal from all of the object all of the time, signal is expected to be high, and each detector is predicted to receive signal of the same order of magnitude as one large detector in conventional MRI. Moreover, the elimination of gradients removes the major source of electrical noise. Unexpectedly, therefore, we anticipate that signal-to-noise (SNR) will be high.

By means of a detailed simulation, in which the exact Hertzian radio field is simulated for each of 32x32x32 voxels interacting with each of 1024 detector loops, we demonstrate that 3-dimensional MR images can indeed be acquired very quickly. Nevertheless, we also simulated imaging in the presence of unforeseen sources of noise, and found that image reconstruction is resistant to noise, even with unrealistically low SNR values.

In summary, high quality 3-Dimensional MRI in millisecond timescales is likely to be feasible.

# Introduction

Lauterbur's revolutionary invention of MRI [1] violated the most fundamental principle of NMR physics: the requirement that the imposed magnetic field, $B_0$, is homogeneous. Because of this homogeneity, NMR signal is dominated by microscopic interactions between nuclear spins and is therefore sensitive to *microscopic* structure. In contrast, MRI renders the applied magnetic field inhomogeneous by deploying magnetic field gradients. Proton spins now experience different field strengths, and therefore different frequencies, depending on their *macroscopic* location. This is gradient encoding of spatial location. While MRI has revolutionized diagnosis, it remains slow because of the requirement of sequential application of hundreds, even thousands, of magnetic field gradients. And although the shortest acquisition time for volumetric imaging has been pared to as little as 100 ms, image quality is low.

In the original embodiment of MRI, three magnetic field gradients were imposed, one for slice selection, one for phase-encoding, one for readout, each applied sequentially. Although there have been several advances in terms of speed, high quality clinical MR imaging still takes minutes.

Mansfield's echoplanar imaging [2] applied phase-encoding and readout gradients simultaneously, reducing imaging time substantially, but remaining prone to artifact and other limitations which preclude its general use.

Imaging speed may also be increased by reducing the number of gradient-encoding steps. This is parallel MRI, [3-10] where the detector is divided into a number of smaller detectors, each having distinct spatial relationships with proton spins in the object. This spatial sensitivity encoding is used to substitute for gradient encoding by permitting the undersampling of k-space. If the degree of undersampling is denoted by the integer M, then imaging is accelerated by a factor of M. Parallel MRI requires N detectors, where $N \geq M$. It has been found empirically that when $M > 4$ the image suddenly degrades. [5-8] This is because the inversion matrices used to construct the image are ill-conditioned, having spatial sensitivity profiles that are not granular, i.e. their profiles vary slowly with space, and this in turn is because the detectors, although smaller, are still large.

With conventional parallel MRI, therefore, spatial sensitivity encoding is less efficient than gradient encoding. When M = 5, 80% of the encoding is performed by spatial sensitivity and only 20% by gradients. Empirically this is unsustainable, and it is now widely accepted that parallel MRI is limited to an M of 4. A combination of multi-slice EPI with parallel MRI (M = 4), yields a net acceleration of 16, and a 3-dimensional image can be made in as little as a few hundred milliseconds, [9] but with the same limitations as EPI. Higher accelerations can be achieved by increasing the number of slice excitations, but although imaging time can be reduced to as little as 100 ms, there is a corresponding reduction in signal-to-noise (SNR). [10] Many other approaches to fast imaging, using various combinations of EPI and parallel MRI, or even eschewing EPI but requiring multiple acquisitions, are beyond the scope of this paper. To our knowledge, what all previous approaches have in common is that they require at least hundreds of milliseconds for volume imaging. This is *entirely* because all previous approaches make the tacit assumption that while gradient encoding can be reduced substantially, it cannot be eliminated.

Nevertheless, it was recently hypothesized that if the number of detectors is greatly increased, and the size of each detector is thereby greatly reduced, the granularity of spatial sensitivity maps would be increased to such an extent that the image would paradoxically be more stable, not less so. [11] This is ULTRA (Unlimited Trains of Radio Acquisitions), which eliminates gradient reversals entirely and encodes solely by means of spatial sensitivity with a single fixed magnetic field gradient. Echoes are now generated spontaneously, without the need for further excitation.

In the current embodiment of ULTRA, N equally spaced small detector loops are arranged around the circumference of a circular ring. N rings are then arranged parallel to one another in the configuration of a cylinder. The z axis is the axis of the cylinder, and a single fixed magnetic gradient is applied along z. The proton spins in the xy plane of the object at position z will therefore all experience the same magnetic field, which will be different from all other planes of the object.

The elimination of gradient reversals confers two massive and unexpected advantages, namely that signal is maintained while noise is greatly reduced. After the initial excitation, echoes form spontaneously with each echo encoding for the entire 3-dimensional image. In removing gradient reversals, eddy currents are eliminated along with just about all electrical noise.

An obvious first objection is that small surface detectors will be less sensitive to deep tissue. As we have shown, however, this is not the case because deep tissue is well represented in the set of all coils, compensating for the fact that its representation in any given coil is relatively small. [11] Conversely, superficial tissue may be highly represented in local coils but poorly represented in distant coils. These disparities in spatial profiles lend a high degree of spatial encoding granularity when coils are numerous and small.

An obvious second objection is that since ULTRA detectors are necessarily very small, the signal in each detector must be correspondingly small. This is also not the case. The embodiment of ULTRA described here deploys N circular rings of N detectors, so that there are $N^2$ detectors. The area of each detector is therefore of order $N^2$ times smaller than a conventional detector, so the signal in each detector is expected to be $N^2$ times smaller than the conventional signal, all else being equal. But all else is not equal because each ULTRA echo contains *all* of the signal from *all* of the object *all* of the time. In conventional MRI, each echo represents a single line in k-space. Since k-space is 3-dimensional, it follows that the amplitude of the conventional MR echo is diminished by a factor of order $N^2$ compared with ULTRA. The small size of the ULTRA detector is therefore fully compensated by the fact that it is receiving signal from the entire 3-dimensional object at all times. In other words, the signal amplitude of an echo in each small ULTRA detector loop is predicted to be of the same magnitude as the amplitude of an echo in conventional MRI in one large detector.

But there is more. Electrical noise is practically eliminated since there are no gradient reversals and no eddy currents. Putting all of this together, it can now be predicted that the signal-to-noise in each ULTRA image is actually *greater* than a conventional MR image, despite its high speed. This is an astonishing result, never anticipated in the first parallel MRI proposals. [3,4]

## Methods

We start with the total signal in each detector d, which can be written as:

$$S^d = \sum_{i,j,k}^{N} \frac{\rho_{i,j,k} \sin(\theta_{i,j,k}^d)}{(r_{i,j,k}^d)^3} e^{i\omega_k t} \qquad (1)$$

where i = 1, N; j = 1,N; k = 1, N. (i,j) labels one voxel in one slice, and k labels the N slices. $\omega_k$ is the frequency of slice k, $\rho_{i,j,k}$ is the spin density in voxel (i,j,k), $\theta_{i,j,k}^d$ is the angle between the dipole source and the line joining the detector and the dipole source, $r^d$ is the distance from a particular voxel (i,j,k) to detector d.

$S^d$ is now Fourier transformed to a power spectrum denoted $A^d$ where there are N amplitudes as a function of $\omega_k$. It is instructive to consider the amplitude of the power spectrum, in detector d, at frequency $\omega_k$. This is just:

$$A_{\omega_k}^d = \sum_{i,j}^{N} \frac{\rho_{i,j,k} \sin(\theta_{i,j,k}^d)}{(r_{i,j,k}^d)^3} \qquad (2)$$

In the present case there are $N^2 \times N^2$ values of $\rho_{i,j,k}$ for slice k, one for each voxel, and all spins in the plane have one frequency, $\omega_k$. All of these spins, taken together, give rise to a signal in each of N detectors. Each detector has one value for the amplitude of the power spectrum at the frequency $\omega_k$. Since there are N detectors there are N values of the amplitude of the power spectrum at frequency $\omega_k$.

The matrix representation of (2) is:

$$(A^d_{\omega_k})_{N^2 x 1} = D_{N^2,N^2} \cdot \rho_{N^2 x 1} \qquad (3)$$

Where D is an $N^2 x N^2$ matrix representing the geometric factor; $\rho$ and $A$ are one dimensional vectors each having $N^2$ elements.

Explicitly this can be written as:

$$\begin{pmatrix} A^1_{\omega_k} \\ A^2_{\omega_k} \\ \vdots \\ \vdots \\ A^{N^2}_{\omega_k} \end{pmatrix} = \begin{pmatrix} \frac{\sin(\theta^1_{1,1,k})}{(r^1_{1,1,k})^3} & \cdots & \frac{\sin(\theta^1_{N,N,k})}{(r^1_{N,N,k})^3} \\ \frac{\sin(\theta^2_{1,1,k})}{(r^2_{1,1,k})^3} & \cdots & \frac{\sin(\theta^2_{N,N,k})}{(r^2_{N,N,k})^3} \\ \vdots & & \\ \frac{\sin(\theta^{N^2}_{1,1,k})}{(r^{N^2}_{1,1,k})^3} & \cdots & \frac{\sin(\theta^{N^2}_{N,N,k})}{(r^{N^2}_{N,N,k})^3} \end{pmatrix} \begin{pmatrix} \rho_{1,1,k} \\ \vdots \\ \vdots \\ \rho_{N,N,k} \end{pmatrix} \qquad (4)$$

An inversion technique is then applied sequentially to obtain each slice of the volume, i.e.

$$\rho_{N^2 x 1} = D^{-1}_{N^2,N^2} \cdot (A^d_{\omega_k})_{N^2 x 1} \qquad (5)$$

$\rho$ is now reformatted from $\rho_{N^2 x 1}$ to $\rho_{N,N}$ which corresponds to the image of spin densities in slice k. This is then solved sequentially for all slices, resulting in the volume image.

The detailed simulation begins with the exact Hertzian radio field at each detector from a phantom of two cylinders, side-by-side, divided into 32x32x32 voxels. Each of the 32,768 voxels in the object contributes to the signal in each of the $N^2$ (=1024) detectors. Each of these signals then undergoes Fourier transformation, yielding 1024 amplitudes at each of 32 frequencies, each frequency corresponding to a particular slice of the object. The 3-dimensional image is then constructed, slice by slice, by inversion.

To make this quite clear, we provide a concrete illustration of the methodology. The matrix Eq. (4), is written out below in Eq. (6) as a set of 1024 equations for the particular case of slice 1, which has frequency $\omega_1$. The amplitude in the first detector, at this frequency, is the first in a set of 1024 equations. In each equation (i.e. each line of Eq. 6), the LHS is the amplitude, in a particular detector, at frequency $\omega_1$, the RHS is the sum over all 1024 pixels in slice 1:

$$A_{\omega_1}^1 = \frac{\rho_{1,1,1}\sin(\theta_{1,1,1}^2)}{(r_{1,1,1}^1)^3} + \frac{\rho_{1,2,1}\sin(\theta_{1,2,1}^1)}{(r_{1,2,1}^1)^3} + ... + \frac{\rho_{2,1,1}\sin(\theta_{2,1,1}^1)}{(r_{2,1,1}^1)^3} + \frac{\rho_{2,2,1}\sin(\theta_{2,2,1}^1)}{(r_{2,2,1}^1)^3} + ... + \frac{\rho_{N,N,1}\sin(\theta_{N,N,1}^1)}{(r_{N,N,1}^1)^3}$$

$$A_{\omega_1}^2 = \frac{\rho_{1,1,1}\sin(\theta_{1,1,1}^2)}{(r_{1,1,1}^2)^3} + \frac{\rho_{1,1,1}\sin(\theta_{1,1,1}^2)}{(r_{1,2,1}^2)^3} + ... + \frac{\rho_{2,1,1}\sin(\theta_{2,1,1}^2)}{(r_{2,1,1}^2)^3} + \frac{\rho_{2,2,1}\sin(\theta_{2,2,1}^2)}{(r_{2,2,1}^2)^3} + ... + \frac{\rho_{N,N,1}\sin(\theta_{N,N,1}^2)}{(r_{N,N,1}^1)^3} \quad (6)$$

$$\vdots$$

$$A_{\omega_1}^{1024} = \frac{\rho_{1,1,1}\sin(\theta_{1,1,1}^{1024})}{(r_{1,1,1}^{1024})^3} + \frac{\rho_{1,2,1}\sin(\theta_{1,2,1}^{1024})}{(r_{1,2,1}^{1024})^3} + ... + \frac{\rho_{2,1,1}\sin(\theta_{2,1,1}^{1024})}{(r_{2,1,1}^{1024})^3} + \frac{\rho_{2,2,1}\sin(\theta_{2,2,1}^{1024})}{(r_{2,2,1}^{1024})^3} + ... + \frac{\rho_{N,N,1}\sin(\theta_{N,N,1}^{1024})}{(r_{N,N,1}^{1024})^3}$$

There are 1024 variables $\rho_{i,j,1}$ in each line of Eq.(6), and 1024 equations for the amplitude at frequency $\omega_1$ in each of 1024 detectors. It is now clear by inspection that there is sufficient information in Eq. (6) to solve, by linear algebraic means, for each value of $\rho$ in slice 1. What is not known is if the matrix D is ill-conditioned or if the inversion process is resistant to unforeseen noise.

The process is repeated for each value of $\omega$, allowing the entire 3-dimensional image to be constructed, slice by slice. A phantom data volume was used with image resolution of NxN = 32x32. The data were generated using a phantom of two cylinders with different diameters.

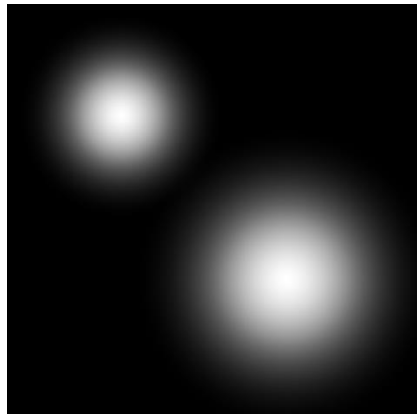

**Figure 1** Cross-section through the original phantom of two cylinders with different diameters, the images of which are shown in Figures 3 and 4.

The z coordinate runs from 0 to 31, z = 0 being the first slice, z = 1 the second, etc. A fixed gradient is imposed along z, so in any given xy slice protons have the same frequency, each slice having a different frequency from all other slices. There are N frequencies. The field of view is FOV=24, bandwidth = 16.384 kHz, divided into 32 frequencies (512 Hz …16384 Hz, in steps of 512 Hz). According to Nyquist's theorem the sampling frequency, $F_S$, must be at least twice the maximum frequency to avoid aliasing. Here $F_S$ is 3 times maximum frequency (i.e. 49152 Hz).

Because there are no gradients, after the initial RF excitation echoes form spontaneously, without further input. The waveform is very different from conventional MRI. Conventional echoes have a much broader width in time. This is consistent with the narrow frequency bandwidth in conventional MRI. With ULTRA, all frequencies in the object are sampled simultaneously, the bandwidth is much wider, and the waveform of the echo is correspondingly much narrower.

Although we assert that electrical noise is practically eliminated by our approach, nevertheless we also considered the possibility that unforeseen noise sources would be present, and simulated the presence of Gaussian noise in each detector, with an unrealistically low SNR of 2.6.

## Results

Three spontaneous echoes forming in a single detector are shown in Figure 2. The ordinate is amplitude and the abscissa is time (in milliseconds).

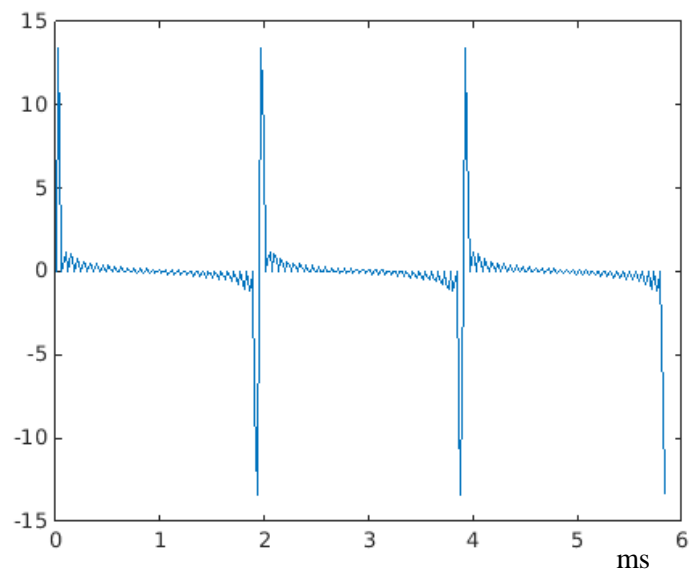

**Figure 2** The waveform in the 16th detector of the 16th detector ring, showing 3 successive, spontaneously forming, echoes without noise. The entire 3-dimensional volume image can be reconstructed from each echo of 2 ms duration.

Note that these echoes are quite different from the echoes seen in conventional MRI. In ULTRA, all of the signal from all of the object is detected all of the time. Therefore, there is a large frequency bandwidth compared with conventional MRI, and a corresponding narrowing of amplitude with respect to time. The reconstructed image, with no noise, is shown in Fig. 3. The reconstructed image, with Gaussian noise in each detector, is shown in Fig. 4.

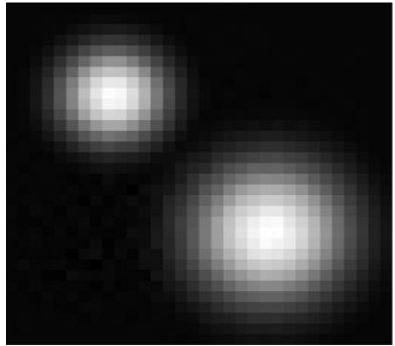

**Figure 3** Slice 16 of the reconstructed image of the phantom, constructed from a single echo, without noise.

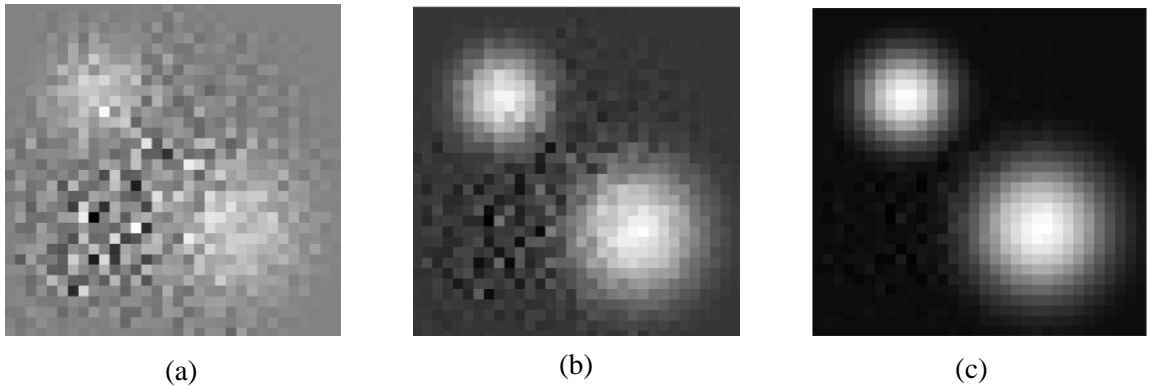

**Figure 4** Simulation of the effects of Gaussian noise on the reconstructed image: a) 2 ms acquisition, SNR = 2.6; b) image at 10 ms; c) image at 20 ms.

# Discussion

We have given a complete mathematical description of the theory of ULTRA MRI, a technique based on the elimination of gradient reversals. To be brief, very large numbers of very small detectors allow for highly granular spatial sensitivity profiles, making spatial sensitivity encoding efficient. This allows elimination of gradient switching. Because there are no gradient reversals, echoes form spontaneously, with each echo coding for the entire volume.

A detailed computer simulation has also been performed, starting with the exact Hertzian radio field for each of 32,768 voxels interacting with each of 1,024 detectors. This demonstrates that 3-dimensional imaging of a phantom can be performed in milliseconds, even in the presence of substantial noise.

However, we also predict that noise levels will be small, signal will be maintained, and the SNR will be high. Relaxation Times $T_1$ and $T_2$ can now be defined by the evolution of echoes over time. Therefore spin density, $T_1$ and $T_2$-weighted images can be constructed in a few hundred milliseconds. Inversion recovery sequences, such as FLAIR, would require relaxation of all spins, so clinical imaging would take longer but would be complete in a few seconds.

Note that sub-millisecond imaging, with the same SNR, might be achieved with higher fixed gradients. Electrical noise is now dominated by detector electronics and coil coupling, and may be further reduced by use of low temperatures and high impedance coils.

Imaging can be performed in milliseconds, therefore neuronal interactions might be detected in real time, since the spin density images include a small BOLD contribution. Functional MRI will therefore be dynamic. Furthermore, since clinical imaging can now be performed in seconds, almost all diagnostic issues of medical significance may in the future be captured by MRI. Finally, total body imaging will now be feasible, suggesting the possibility of routine, computer driven, early detection of cancer.